\documentclass[aps,reprint,twocolumn,superscriptaddress,citeautoscript,showpacs,amsmath,amssymb,amsfonts,floatfix,prb]{revtex4-1}

\usepackage{graphicx}% Include figure files
\usepackage{dcolumn}% Align table columns on decimal point
\usepackage{bm}% bold math
\usepackage{times}% font that prx use
\usepackage{color}
\begin{document}
\title{Short-Range Magnetic Correlations in the Highly-Correlated Electron Compound CeCu$_{4}$Ga}

\author{B. G. Ueland}
\email{bgueland@ameslab.gov;  bgueland@gmail.com}
\affiliation{Los Alamos National Laboratory, Los Alamos, NM 87544, USA}
\affiliation{Ames Laboratory, U.S. DOE, Iowa State University, Ames, Iowa 50011, USA}
\affiliation{Department of Physics and Astronomy, Iowa State University, Ames, Iowa 50011, USA}

\author{C. F. Miclea}
\affiliation{Los Alamos National Laboratory, Los Alamos, NM 87544, USA}
\affiliation{National Institute of Materials Physics 077125 Bucharest-Magurele, Romania}

\author{K. Gofryk}
\affiliation{Los Alamos National Laboratory, Los Alamos, NM 87544, USA}
\affiliation{Idaho National Laboratory, Idaho Falls, Idaho 83415, USA}

\author{Y. Qiu}
\affiliation{NIST Center for Neutron Research, National Institute of Standards and Technology, Gaithersburg, MD 20899, USA}
\affiliation{Department of Materials Science and Engineering, University of Maryland, College Park, MD 20742, USA}

\author{F. Ronning}
\affiliation{Los Alamos National Laboratory, Los Alamos, NM 87544, USA}

\author{R. Movshovich}
\affiliation{Los Alamos National Laboratory, Los Alamos, NM 87544, USA}

\author{E. D. Bauer}
\affiliation{Los Alamos National Laboratory, Los Alamos, NM 87544, USA}

\author{J. S. Gardner}
\affiliation{National Synchrotron Radiation Research Center, Neutron Group, Hsinchu 30077, Taiwan}

\author{J. D. Thompson}
\affiliation{Los Alamos National Laboratory, Los Alamos, NM 87544, USA}

\date{\today}

\begin{abstract}
We present experimental results for the heavy-electron compound CeCu$_{4}$Ga which show that it possesses short-range magnetic correlations down to a temperature of $T = 0.1$~K.  Our neutron scattering data show no evidence of long-range magnetic order occurring despite a peak in the specific heat at $T^{*} =1.2$~K.  Rather, magnetic diffuse scattering occurs which corresponds to short-range magnetic correlations occurring across two unit cells.   The specific heat remains large as $T\sim0$~K resulting in a Sommerfeld coefficient of $\gamma_{0} = 1.44(2)$~J/mol-K$^{2}$, and, below $T^{*}$, the resistivity follows $T^{2}$ behavior and the ac magnetic susceptibility becomes temperature independent. A magnetic peak centered at an energy transfer of $E_{\rm{c}}=0.24(1)$~meV is seen in inelastic neutron scattering data which shifts to higher energies and broadens under a magnetic field.  We discuss the coexistence of large specific heat, magnetic fluctuations, and short-range magnetic correlations at low temperatures and compare our results to those for materials possessing spin-liquid behavior.
\end{abstract}

\maketitle

Highly-correlated electron metals characteristically host multiple competing electronic interactions of comparable strengths that lead to closely spaced low-energy macroscopic states.  This competition often yields novel phenomena such as unconventional superconductivity, unconventional magnetic phase transitions, quantum critical behavior, and strong magnetic spatial correlations in the absence of long-range order \cite{Stewart_1984,Coleman_2007,Si_2010, Balents_2010}.  In particular, heavy-fermion metals are highly-correlated electron materials possessing itinerant charge mediated Ruderman-Kittel-Kasuya-Yosida (RKKY) magnetic exchange, as well as antiferromagnetic (AFM) Kondo-coupling between their localized spins (total electronic magnetic moments) and the spins of their itinerant charges. When these two interactions have comparable strengths, the specific heat becomes large as the temperature goes to $T\sim0$~K reflecting a correspondingly large low temperature entropy arising from competing low energy states and associated fluctuations.
	
CeCu$_5$ is the parent member of the series of highly-correlated electron compounds CeCu$_{\textrm{5-}x}M_{x}$, $x = 0, 1, 2,~M =$ Ga, Al, which are hexagonal metals described by the space group $P6/mmm$, and  possess magnetic Ce atoms which lie in the basal plane and form a potentially geometrically frustrated magnetic sublattice of side-sharing triangles \cite{Bauer_1987,Diep_2004}.  There are two distinct Cu sites: the 2c site in the plane of Ce atoms and the 3g site lying half a lattice constant above the basal plane.  CeCu$_5$ possesses AFM order below $T_{\rm{N}} = 3.9$~K with a magnetic propagation vector of $\bm{\tau}=(0,0,\frac{1}{2})$, an ordered moment of $0.36(4)\mu_{\rm{B}}$, and spins that order collinear with the crystalline $c$-axis \cite{Willis_1987,Bauer_1994}.

CeCu$_{4}$Ga is a member of the series that possesses no long-range magnetic order down to at least $T = 0.03$~K despite peaks in its specific heat and magnetic susceptibility at $T^{*} = 1.2$~K \cite{Wiesinger_1997,Bauer_1988,Kohlmann_1989}. The Ga in the material substitutes preferentially and randomly for the Cu at 3g sites, with little disorder in the plane of Ce atoms, \cite{Kim_1991,Moze_1995} and the crystal field environment of the Ce spins should split the putative $J = \frac{5}{2}$ single-ion ground state multiplet into three magnetic doublets.  Previous neutron scattering experiments have found that a crystal field level transition occurs for a neutron energy transfer of $E=6$~meV, and that the transition is broadened due to Kondo-coupling \cite{Gignoux_1990}.  In addition, CeCu$_{4}$Ga possesses a  large Sommerfeld coefficient \cite{Bauer_1988,Kohlmann_1988} (over 1~J/mol-K$^{2}$)  which could be interpreted as being due to the presence of itinerant charge carriers with an effective mass hundreds of times greater than that of a free electron \cite{Coleman_2007,Stewart_1984,Thompson_2000,Thompson_1994}.  Our data, however, suggest an alternative interpretation for this large $\gamma_{0}$.  

Here, we present results from experiments on CeCu$_{4}$Ga which show that it possesses short-range magnetic correlations at $T = 0.1$~K.  A peak occurs in its specific heat at $T^{*} =1.2$~K, below which its ac magnetic susceptibility becomes temperature independent and its resistivity appears to follow $T^{2}$ behavior.  Our neutron scattering data taken down to $T=0.1$~K do not show any magnetic Bragg peaks, which would correspond to long-range magnetic order, but show a modulation in neutron momentum transfer $Q$ that corresponds to short-range magnetic correlations occurring across two unit cells.  We compare our data to those for materials possessing spin-liquid behavior.

Polycrystalline samples of CeCu$_{4}$Ga and LaCu$_{5}$ were prepared by arc melting the constituents on a water cooled Cu hearth under an ultra-high purity (99.999\%) Ar atmosphere, and were determined to be single phase by powder x-ray diffraction.  A single crystal of CeCu$_{4}$Ga was grown by the Czochralski method and oriented using x-ray Laue backscattering.  The magnetization $M$ was measured down to $T=1.8$~K and in magnetic fields up to $\mu_{0}H = 5.5$~T in a Quantum Design Superconducting Quantum Interference Device (SQUID) to obtain the dc magnetic susceptibility $\chi=\frac{M}{\mu_{0}H}$.  A Quantum Design Physical Properties Measurement System (PPMS) was used to measure the longitudinal resistivity $\rho_{\rm{xx}}$  and specific heat down to $T = 0.45$~K and in fields up to $\mu_{0}H = 9$~T.  Specific heat measurements were continued down to $T = 0.05$~K in a $^{3}$He/$^{4}$He dilution refrigerator using a semi-adiabatic heat pulse technique.  The magnetic specific heat $C_{\rm{mag}}$ was determined by measuring and subtracting off the specific heat of LaCu$_{5}$, which is a non-magnetic isostructural analogue of CeCu$_{4}$Ga with a presumably similar phonon spectrum.  The low temperature nuclear contribution to the specific heat has also been subtracted.   Measurements of the ac magnetic susceptibility $\chi_{\rm{ac}}$ were made down to $T = 0.35$~K and in fields up to $\mu_{0}H = 1$~T using a mutual inductance bridge anchored to a $^{3}$He refrigerator.  A sinusoidally oscillating field of $\mu_{0}H_{\rm{ac}}\approx 10^{-4}$~T was applied by a superconducting primary coil  at frequencies spanning $f = 0.2-5$~kHz.  Inelastic neutron scattering experiments were performed on polycrystalline samples with the Disc Chopper Spectrometer (DCS) \cite{Copley_2003} at the NIST Center for Neutron Research using neutrons with incident wavelengths of $\lambda = 4.8$ or 1.8~\AA.  The sample was placed in a Cu can containing He exchange gas and cooled in a dilution refrigerator down to $T = 0.1$~K.  Fields up to $\mu_{0}H = 5$~T were applied, and counting times ranged from $2-6$ hours per spectrum.  The DCS data  have been multiplied by $\bigl(\frac{k_{\rm{i}}}{k_{\rm{f}}}\bigr)^{4}$ and normalized to the incident beam monitor.

%Figure 1
\begin{figure}[]
\centering
\includegraphics[width=1.0\linewidth]{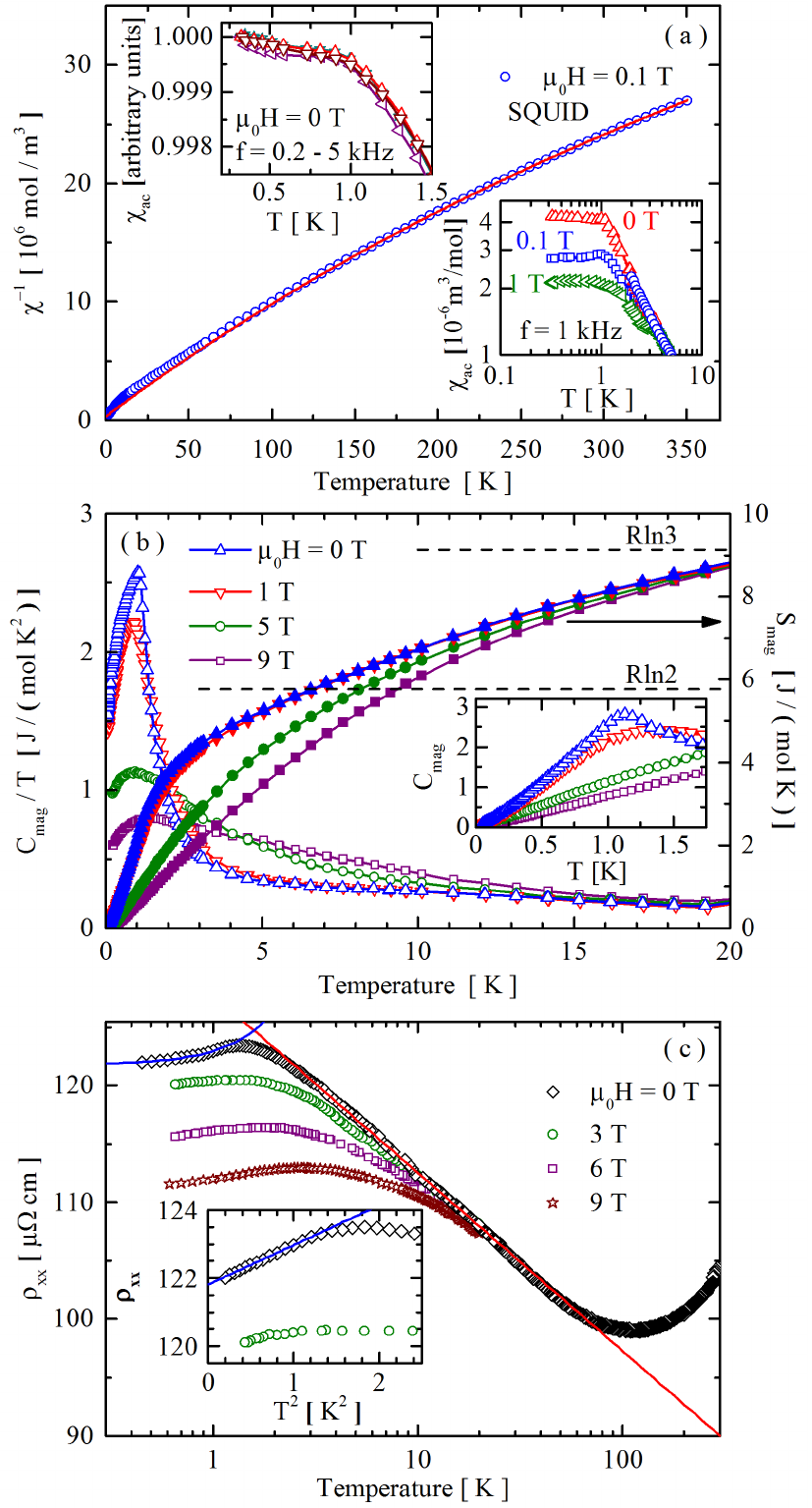}
\caption{(Color online)  (a) $\chi^{\textrm{-1}}$ for $\mu_{0}H=0.1$~T.  The line is a fit as described in the text.  (top inset)  $\chi_{\rm{ac}}$ for $f=0.1$, 0.5, 1, and 5~kHz.  (bottom inset)  $\chi_{\rm{ac}}$ for $f=1$~kHz and various applied dc fields after normalizing to the dc susceptibility data also shown (blue circles).  (b) $C_{\rm{mag}}/T$ (left axis) and the magnetic entropy (right axis) for various fields.  (inset) $C_{mag}$ for various fields.  (c)  $\rho_{\rm{xx}}$ for various fields.  (inset) $\rho_{\rm{xx}}$  versus $T^{2}$.  The blue lines represent a fit to $\rho_{\rm{xx}}=\rho_{0}+AT^{2}$.  The red line is a fit showing that $\rho_{\rm{xx}}\propto-\ln T$ for $T=5$~to 50~K.  Uncertainties are within the size of the symbols unless otherwise indicated and are statistical in origin, representing one standard deviation. \label{Fig1}}
\end{figure}

Figure~\ref{Fig1}a shows the temperature dependence of $\chi^{\textrm{-1}}$ of a polycrystalline sample with $\mu_{0}H = 0.1$~T.  The line through the data is a fit between $T = 250$~to 350~K which assumes that $\chi$ is the sum of the Curie-Weiss and temperature independent susceptibilities \cite{Ashcroft_1976}.  The fit yields a Weiss-temperature of $\theta_{\rm{W}} = -3(1)$~K, which indicates that the effective interaction between spins is weakly AFM, an effective moment of $p_{\rm{eff}}= 2.45(6)~\mu_{\rm{B}}$, which is slightly lower than the expected value of $p_{\rm{eff}}=2.54~\mu_{\rm{B}}$ possibly due to crystal-field or Kondo-coupling effects, and a temperature independent susceptibility of $\chi_{\rm{0}}=4.23(5)~10^{-6}\rm{m}^{3}$/mol.  The fit deviates from the data below $T\approx 150$~K.  The bottom inset shows $\chi_{\rm{ac}}$ for a single crystal sample measured in a $f=1$~kHz ac field while applying various static fields.  In zero field, $\chi_{\rm{ac}}$ monotonically increases with decreasing temperature down to $T^{*}=1.2$~K, below which it is practically temperature independent.  Application of a $\mu_{0}H=1$~T field suppresses $\chi_{\rm{ac}}$, but its affect on $T^{*}$ is not clear.  For $\mu_{0}H=0$~T, $\chi_{\rm{ac}}(T\sim0~\rm{K})=4.311(1)~10^{-6}\rm{m}^{3}$/mol which is close to the value determined for $\chi_{\rm{0}}$.  The top inset shows that $T^{*}$ does not change for different measurement frequencies, which indicates that spin freezing does not occur on the time scale of the measurement.  Though not shown, we also measured $M(\mu_{0}H)$ at $T=2$~K and found $M=0.733~\mu_{\rm{B}}$/Ce at $\mu_{0}H=6$~T which is consistent with previous magnetization results assigning the Ce ground state as a $|J_{z}|=\frac{1}{2}$ magnetic doublet \cite{Bauer_1988}.

Figure~\ref{Fig1}b is a plot of $C_{\rm{mag}}/T$ (left axis) and the integrated magnetic entropy $S_{\rm{mag}}$ (right axis) for a single crystal sample, and the inset shows $C_{\rm{mag}}$ at low temperatures.  A peak in $C_{\rm{mag}}/T$ occurs at $T^{*}$ which decreases, broadens, and shifts to higher temperature with increasing field.  Extrapolating the zero field data below $T^{*}$ to $T=0$~K yields $\gamma_{0}=1.44(2)$~J/mol-K$^2$, which is indicative of the presence of  heavy itinerant charge carriers and/or low-energy magnetic excitations.  $C_{\rm{mag}}$ cannot be fit to a Schottky-term which suggests that the peak at $T^{*}$ does not represent a canonical crystal-field level transition.  Furthermore, $S_{\rm{mag}}$ for $\mu_{0}H=0$~T does not reach the value expected for a ground state doublet of $S_{\rm{mag}}=R\ln2$~J/mol-K until $T\approx6.6$~K.  As we will discuss, our neutron scattering data demonstrate that a broad low-energy magnetic excitation exists both above and below $T^{*}$, centered at an energy comparable to $k_{\rm{B}}T^{*}$.

Figure~\ref{Fig1}c gives $\rho_{\rm{xx}}$ at $\mu_{0}H=0$, 3, 6, and 9~T for a polycrystalline sample.  In zero field, a minimum occurs at $T=115$~K which is indicative of Kondo scattering, followed by a maximum slightly above $T^{*}$ at $T=1.35$~K below which  $\rho_{\rm{xx}}$ decreases.  The red line is a fit showing $\rho_{\rm{xx}}\propto-\ln T$ and indicates that $\rho_{\rm{xx}}$ is dominated by Kondo scattering for $5<T<50$~K.  The blue lines in the main panel and the inset show a fit to $\rho_{\rm{xx}}=\rho_{0}+AT^{2}$ below $T=1.1$~K which may indicate Fermi-liquid type electrical transport.  The fit yields $A=1.16(1)~\mu\Omega\,\rm{cm/K}^{2}$ and $\rho_{0}=121.8(1)~\mu\Omega\,\rm{cm}$.  The very large residual resistivity, $\rho_{0}$,  likely is dominated by scattering induced by the random distribution of Ga on the 3g Cu sites \cite{Kim_1991,Moze_1995}.  With increasing field, the temperature of the maximum in $\rho_{\rm{xx}}$ increases while $|\rho_{\rm{xx}}|$ decreases.  

%Figure 2
\begin{figure}[]
\centering
\includegraphics[width=1.0\linewidth]{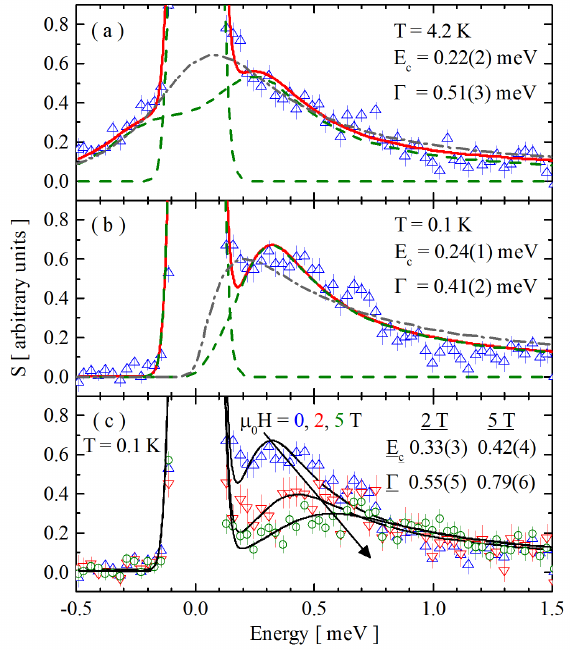}
\caption{(Color online)  The neutron scattering function $S(E)$ obtained by integrating over $Q=0.46$ to 0.56~\AA$^{-1}$ for (a) $T=4.2$~K and (b) 0.1~K.  The solid red lines are the total fits to Eq. \ref{Eq1}, while the dashed green lines show the Gaussian and Lorentzian components of the fits. The dashed-dotted grey lines show the quasielastic components to the fits if the sum of a Gaussian and a quasielastic line-shape is assumed.  (c) $S(E)$ at $T=0.1$~K for various applied magnetic fields obtained by integrating over $Q=0.46$~to 0.56~\AA$^{-1}$.  The black lines are fits to Eq. \ref{Eq1}.  Each panel shows the values for $E_{\rm{c}}$ and $\Gamma$ determined from the fits.  \label{Fig2}}
\end{figure}

Figures~\ref{Fig2}a and~\ref{Fig2}b are plots of the neutron scattering function $S$ as a function of $E$ for $T=4.2$ and 0.1~K, respectively, obtained by integrating DCS data for $\lambda=4.8$~\AA~incident neutrons over $Q=0.46-0.56$~\AA$^{-1}$. The solid lines are fits to:
\begin{eqnarray} 
	S(E)=\frac{A}{w\sqrt{\frac{\pi}{4\ln2}}}e^{-\frac{E^{2}4\ln2}{w^{2}}}+\Biggl(\frac{E}{ 1-e^{-\frac{E}{k_{\rm{B}}T}}} \Biggr) \label{Eq1} \\* \nonumber  \times \frac{2B\Gamma}{\pi} \Biggl[ \frac{1}{4(E-E_{\rm{c}})^{2}+\Gamma^{2}} +\frac{1}{4(E+E_{\rm{c}})^{2}+\Gamma^{2}} \Biggr].
\end{eqnarray}    
The first term is a Gaussian line-shape that describes mainly the incoherent scattering centered at $E=0$~meV, and the fits yield a full width at half maximum of $w=0.11(1)$~meV for both temperatures which corresponds to the expected experimental resolution. $A$ is the area.  The remaining terms are modified Lorentzian line-shapes which describe scattering due to excitations centered at $E_{\rm{c}}$ multiplied by a detailed-balance factor \cite{Lovesey_1984}. $B$ is proportional to the area of each Lorentzian line-shape and $\Gamma$ is the full width at half maximum.  For both temperatures the values of $E_{\rm{c}}$ and $\Gamma$ are constant within error over $Q=0.2$~to 2.2~\AA$^{-1}$, and $B$ decreases with increasing $Q$, which is consistent with the excitations being magnetic.  We also have performed fits assuming that the data in Fig. 2 may be described by the sum of a Gaussian line-shape and a quasielastic line-shape given by $(2B\Gamma E / \pi) (4E^{2}+\Gamma^{2})^{-1}$.  The resulting quasielastic components are shown by the dashed-dotted grey lines in Figs. \ref{Fig2}a and  \ref{Fig2}b.  While the fit may sufficiently describe the $T=4.2$~K data, it underestimates the intensity around $E=0.5$~meV for $T=0.1$~K.  Further measurements are necessary to conclusively determine the best fit, but here we assume that Eq. \ref{Eq1} models the data. 

Figure~\ref{Fig2}c shows that for $T=0.1$~K application of a magnetic field suppresses the inelastic peak, which is also indicative of magnetic scattering.  Qualitatively, the field induced changes to the peak appear to reflect the field dependent evolution of the low temperature specific heat and resistivity, suggesting that each may have a common origin.  One possibility is that the scattering peak corresponds to a low-lying crystal field level broadened by Kondo scattering.  However, this should result in an entropy of $S_{\rm{mag}}=R\ln4$~J/mol-K at a temperature corresponding to a few times $E_{\rm{c}}$, which we do not observe. Alternatively, as discussed below, the diffuse scattering data indicate the presence of additional magnetic scattering at low temperatures.

%Figure 3
\begin{figure}[]
\centering
\includegraphics[width=1.0\linewidth]{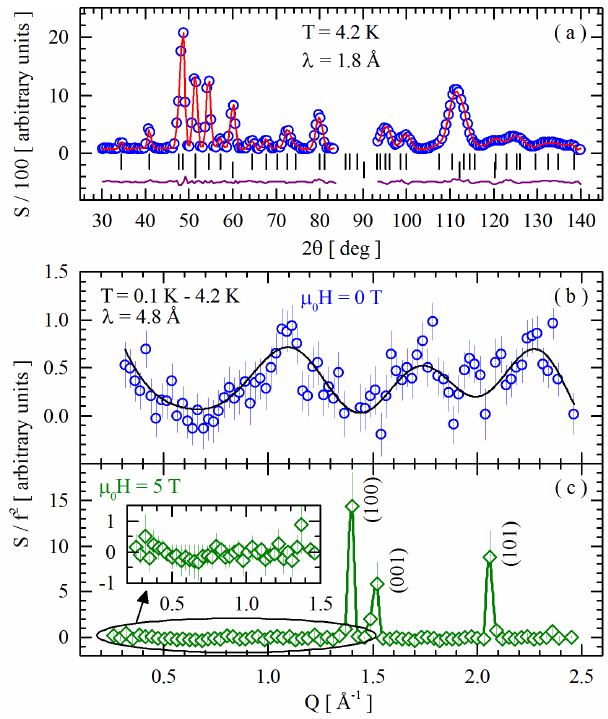}
\caption{(Color online)  (a) $T=4.2$~K diffraction pattern obtained using $\lambda=1.8$~\AA~incident neutrons and integrating over $E=-2$ to 2~meV data.  The line through the data is a Reitveld refinement, vertical lines indicate Bragg peak positions for the sample (upper) and the Cu sample can (lower), and the bottom line is the difference between the data and fit.  Data are masked out in the middle of the figure due to the Cu sample can possessing some preferred orientation.  (b) $T=0.1$~K diffuse scattering data for $\lambda=4.8$~\AA~incident neutrons after integrating over $E=-0.1$ to 0.1~meV and subtracting by the corresponding $T=4.2$~K data.  The data have also been divided by the square of the Ce$^{3+}$ magnetic form factor $f$.  The line is a fit to Eq. \ref{Eq2}.  (c) Diffuse neutron scattering data for $\mu_{0}H= 5$~T constructed as in (b).  The field-induced peaks are indexed according to the chemical lattice.  The inset shows a blow-up of the low $Q$ data. \label{Fig3}}
\end{figure}

Figure~\ref{Fig3}a plots the diffraction pattern at $T=4.2$~K for $\lambda=1.8$~\AA~incident neutron data constructed by integrating over $E=-2$~to 2~meV.  The line through the data is a fit from a Rietveld refinement performed using  \textsc{fullprof} \cite{Rodriguez-Carvajal_1993}, and has a goodness of fit parameter $R_{\rm{Bragg}}=2.70$\% which indicates that the sample possesses the anticipated structure.  To check for the existence of magnetic Bragg peaks below $T^{*}$, which would indicate long-range magnetic order, $T=0.1$ and 4.2~K data taken using  $\lambda=4.8$~\AA~incident neutrons were integrated over $E=-0.1$~to 0.1~meV to construct diffraction patterns.  The $T=0.1$~K pattern was then subtracted by the $T=4.2$~K pattern, and the result was divided by the square of the Ce$^{3+}$ magnetic form factor $f$.  The resulting data are plotted in Fig.~\ref{Fig3}b.  No magnetic Bragg peaks were found, however, magnetic diffuse scattering is present in the form of broad peaks centered at $Q\approx1.1$, 1.75, and 2.3~\AA$^{-1}$. 

The diffuse scattering data in Fig.~\ref{Fig3}b are fit to a function describing scattering from isotropic short-range magnetic correlations \cite{Bertaut_1967}:
\begin{equation}
\frac{S(Q)}{f^{2}}=b+\sum_{i}C_{i}\frac{\sin{Qr_{i}}}{Qr_{i}}.
\label{Eq2}
\end{equation}
Here, $i$ indexes correlated pairs of Ce spins, $b$ is the background (a constant offset), $C_{i}$ is a constant determined from the fit, and $r_{i}$ is the distance between two correlated Ce spins.  An initial fit to Eq.~\ref{Eq2} in which $r_{i}$ was allowed to vary found that a minimum of two different $r_{i}$ values are necessary to fit the data: $r_{1}=10.0(5)$~\AA~and $r_{2}=8.5(4)$~\AA.  These values are close to double the values of the $a$ and $c$ lattice parameters, respectively.  Using this information, we again fit the data and included all Ce-Ce distances within two unit cells.  This fit is shown by the black curve, and it reproduces the maxima and minima in the data as well as their widths.  Figure \ref{Fig3}c shows similar data for $\mu_{0}H=5$~T.  The field suppresses the diffuse scattering and induces sharp peaks at $Q$ positions which are commensurate with the chemical lattice.  These results are consistent with the diffuse scattering having a magnetic origin, and the sharp peaks likely occur due to the field polarizing the Ce spins.  By comparing the data in Fig.~\ref{Fig3}c with magnetization versus field data and calculations of hypothetical magnetic structure factors, we estimate that any long-range magnetic order present at $\mu_{0}H=0$~T must possess an ordered moment less than $0.3~\mu_{\rm{B}}$.  We note that muon spin relaxation experiments also found that no long-range order exists down to $T=0.03$~K \cite{Wiesinger_1997}.

While the $\rho_{\rm{xx}}(T)$ data may indicate a crossover from incoherent Kondo scattering dominated electrical transport to possible Fermi-liquid behavior below $T^{*}$, the recovery of entropy to $S_{\rm{mag}}=R\ln2$ at $T>T^{*}$ likely is not due to the onset of coherence sometimes observed in heavy-electron metals \cite{Yang_2008}, since previous experiments found that applying pressure does not lead to a large change in $T^{*}$ \cite{Eichler_1995}. In addition, since the maximum in $\chi_{\rm{ac}}$ persists at $\mu_{0}H=1$~T and $T^{*}$ is independent of the frequency of the applied field, it is likely that a spin glass does not form upon cooling through $T^{*}$ \cite{Mydosh_1993}.  Nevertheless, the change in behavior of $\rho_{\rm{xx}}$ at $T^{*}$ signals a switch in the magnetic scattering of the itinerant charges.  The Wilson ratio $R=\frac{\pi^{2}k_{\rm{B}}^{2}\chi_{0}}{\mu_{\rm{B}}^{2}p_{\rm{eff}}^{2}\gamma_{0}}$ is $R=2$ for the spin $\frac{1}{2}$ Kondo-impurity model and is typically $\agt0.8$ for heavy-fermion compounds possessing long-range magnetic order \cite{Stewart_1984}.  However, we calculate $R=8.5(2)$, which suggests, that CeCu$_{4}$Ga should possess magnetic order and ferromagnetic correlations \cite{Aside1}. 

The presence of spin-orbit coupling can modify the Wilson ratio, but the ratio is also large for some candidate quantum spin-liquids \cite{Balents_2010}.  This fact is intriguing since while short-range magnetic correlations have been observed previously for certain heavy-fermion compounds (e.g. CeNi$_{2}$Ge$_{2}$ \cite{Fak_2000}, CeRu$_{2}$Si$_{2}$ \cite{Kadowaki_2004}, and UCu$_{4}$Pd \cite{Aronson_2001}), our data are also reminiscent of those for geometrically frustrated insulators that show large moment spin-liquid behavior (e.g.  Tb$_{2}$Mo$_{2}$O$_{7}$ and Tb$_{2}$Ti$_{2}$O$_{7}$) \cite{Gardner_2010}, as well as  the magnetically frustrated heavy-fermion LiV$_{2}$O$_{4}$ \cite{Kondo_1997,Lee_2001, Krimmel_2003}.  Also, the existence of relatively broad inelastic magnetic scattering and the absence of spin-freezing suggest that the magnetic correlations in CeCu$_4$Ga may be dynamic, which is consistent with muon spin relaxation results that suggest a quasi-static moment exists at low temperatures \cite{Wiesinger_1997}.  Finally, since the modulation in $Q$ of the diffuse scattering does not coincide with CeCu$_{5}$'s AFM propagation vector, future measurements should focus on how the Ga doping suppresses AFM order.  While the origin of the unconventional behaviors in CeCu$_4$Ga currently is unknown, our results point towards future experiments on the CeCu$_{\textrm{5-}x}$Ga$_{x}$ series that explore the mechanism(s) responsible for the observed short-range magnetic correlations and, more generally, how spin correlations and fluctuations manifest in heavy-electron metals with a potentially frustrated magnetic sublattice.  

\begin{acknowledgments}
We are grateful for discussions with and assistance from J. M. Lawrence, I. Martin, C. D. Batista, V. Zapf, R. J. McQueeney, G. S. Tucker, A. Kreyssig, A. I. Goldman, and J. Scherschligt.  Work at Los Alamos National Laboratory was conducted under the auspices of the U.S. Department of Energy and supported in part by the Laboratory Directed Research and Development program.  Work at the Ames Laboratory was supported by the Department of Energy, Basic Energy Sciences, Division of Materials Sciences \& Engineering, under Contract No. DE-AC02-07CH11358.  C. F. M. acknowledges PN-II-ID-PCE-2011-3-1028.  We acknowledge the support of the National Institute of Standards and Technology, U.S. Department of Commerce, in providing the neutron research facilities used in this work, which are supported in part by the National Science Foundation under Agreement No. DMR-0944772.  Certain commercial equipment is identified in this paper to foster understanding. Such identification does not imply recommendation or endorsement by the National Institute of Standards and Technology, nor does it imply that the equipment identified is necessarily the best available for the purpose.
\end{acknowledgments}


\begin{thebibliography}{99}
\bibitem{Aside1} In Ref.\onlinecite{Kohlmann_1989} the Wilson ratio is reported as being $R\approx3.9$, however, since this value is also $\gg 0.8$, it still qualitatively agrees with our discussion.
\end{thebibliography}
\end{document}